\documentclass{lhep}

\journal{LHEP}
\vol{3}      
\jyear{2019}
\pages{9}    

\received{xx January 2019}
\published{xx March 2019}

\usepackage{amsmath,amssymb,revsymb,graphicx}

\begin{document}

\title{Asymmetric nonsingular bounce from a dynamic scalar field}

\author{F.R. Klinkhamer,\auno{1} Z.L. Wang\auno{1}
}
\address{$^1$Institute for Theoretical Physics,
Karlsruhe Institute of Technology (KIT), 76128 Karlsruhe, Germany}

\begin{abstract}
We present a dynamical model for
a time-asymmetric nonsingular bounce with
a post-bounce change of the effective equation-of-state parameter.
Specifically, we consider
a scalar-field model with a time-reversal-noninvariant
effective potential. \hfill  [arXiv:1904.09961]
\end{abstract}

\maketitle


\section{Introduction}
\label{sec:Intro}

A nonsingular bouncing cosmology can have several interesting properties
which may remove the need for inflation;  see the discussion of   
Ref.~\cite{IjjasSteinhardt2018} and references therein.
The crucial question, then, is what physics is responsible for such
a nonsingular bounce of the cosmic scale factor.

Recently, a very simple suggestion has been put forward, namely
to keep the structure of general relativity but to allow for degenerate
metrics [$\det g_{\mu\nu}(x)=0$ at certain spacetime points].
In that case, it is possible to make an  \textit{Ansatz}
for the metric which leads to a modified
Friedmann equation with a nonsingular bouncing
solution~\cite{Klinkhamer2019}.

A follow-up paper~\cite{KlinkhamerWang2019}
has calculated certain cosmological observables
(for the moment, only as \textit{Gedankenexperiments}).
That follow-up paper also
presented, in its Appendix~A, an explicit model for
the time-asymmetric nonsingular bouncing cosmology
discussed in Ref.~\cite{IjjasSteinhardt2018}.
This explicit model was constructed at the hydrodynamic
level with a designer equation-of-state parameter $w(T)$,
where $T$ is the cosmic time coordinate used for the metric
(see Sec.~\ref{subsec:Metric-Ansatz}).

The goal, here, is to obtain a dynamic realization of
a time-asymmetric nonsingular bounce with
a post-bounce change of the effective equation-of-state parameter:
$w_\text{eff,\,pre-bounce}(T)=1$ for $T \leq 0$
and $w_\text{eff,\,post-bounce}(T)<1$ for $T > 0$.

\section{Model with a massive scalar}
\label{sec:Model-with-massive-scalar}

We take a scalar field $\phi(x)$ minimally coupled to
Einstein gravity. The scalar has self-interactions determined
by a special effective potential $V_\text{eff}(\phi)$, which is
possibly related to a fundamental time
asymmetry~\cite{Klinkhamer2002,Klinkhamer2000,Klinkhamer2018},
as will be explained in Sec.~\ref{sec:Discussion}.
Specifically, we consider a homogenous
scalar field $\phi$ that propagates over the spacetime manifold
from Ref.~\cite{Klinkhamer2019}.

In Sec.~\ref{subsec:Metric-Ansatz}, we briefly review the
metric \textit{Ansatz} from Ref.~\cite{Klinkhamer2019}, which
applies to the case of a spatially flat universe.
In Sec.~\ref{subsec:Reduced-field-equations-with-dynamic-scalar-field},
we introduce a dynamic scalar field and consider the
reduced field equations from a particular model.
In Sec.~\ref{subsec:Numerical-analytic-results-model-phi}, we obtain
the numerical solution of these reduced field equations,
together with analytic results for the pre-bounce behavior
of the solution and the asymptotic post-bounce behavior.
Natural units with $c=1$ and $\hbar=1$ are used initially.

\subsection{Metric Ansatz}
\label{subsec:Metric-Ansatz}

With a cosmic time coordinate $T$ and co-moving  
spatial Cartesian coordinates $\{x^{1},\,  x^{2},\, x^{3}\}$,
the metric \textit{Ansatz} for a spatially flat universe
is given by~\cite{Klinkhamer2019}%
\begin{subequations}\label{eq:mod-FLRW}
\begin{eqnarray}\label{eq:mod-FLRW-ds2}
\hspace*{-0mm}
ds^{2}
&\hspace{-2mm}\equiv\hspace{-2mm}&
g_{\mu\nu}(x)\, dx^\mu\,dx^\nu
\nonumber\\[1mm] &\hspace{-2mm}=\hspace{-2mm}&
- \frac{T^{2}}{b^{2}+T^{2}}\,dT^{2}
+ a^{2}(T)
\;\delta_{kl}\,dx^k\,dx^l\,,
\\[2mm]
\hspace*{-0mm}
b &\hspace{-2mm}>\hspace{-2mm}& 0\,,
\\[2mm]
\hspace*{-0mm}
T &\hspace{-2mm}\in\hspace{-2mm}& (-\infty,\,\infty)\,,
\quad
x^k \;\in\; (-\infty,\,\infty)\,,
\end{eqnarray}
\end{subequations}
where $b$ corresponds to the regularization parameter.
The corresponding spacetime manifold has topology $\mathbb{R}^4$.

Observe that the metric \eqref{eq:mod-FLRW-ds2} is degenerate:
$\det\,g_{\mu\nu} = 0$ at $T=0$.
The corresponding spacetime slice at $T=0$ may be
interpreted as a 3-dimensional ``defect'' of spacetime with
topology $\mathbb{R}^3$. The parameter $b$ then corresponds to the
characteristic length scale of this spacetime defect.

Assuming, for the moment, that the matter
content of the cosmological model
is solely given by a homogeneous perfect fluid  
with a relativistic-matter equation of state $P(\rho)=\rho/3$,
the Einstein equation
from the metric \eqref{eq:mod-FLRW-ds2} gives
the following bounce solution of the scale factor~\cite{Klinkhamer2019}:
\begin{equation}\label{eq:bouncing-cosmology-asol}
a(T)=
\left(\frac{b^{2}+T^{2}}{b^{2}+T_{0}^{2}}\right)^{1/4}\,,
\end{equation}
where $a(T)$ has been normalized to unity at $T=T_{0}>0$.
The solution \eqref{eq:bouncing-cosmology-asol}
is perfectly smooth at $T=0$, as long as
$b\neq 0$. The corresponding Kretschmann curvature scalar
$K \equiv R^{\mu\nu\rho\sigma}\,R_{\mu\nu\rho\sigma}$
and the energy density $\rho$ are given by~\cite{Klinkhamer2019}
\begin{subequations}\label{eq:bouncing-cosmology-Ksol-rhosol}
\begin{eqnarray}
K(T) &\hspace{-2mm}=\hspace{-2mm}&  
\frac{3}{2}\;\frac{1}{\big(b^{2}+T^{2}\big)^{2}}\,,
\\[2mm]
\rho(T) &\hspace{-2mm}=\hspace{-2mm}&
\rho_{0}\;\frac{b^{2}+ T_{0}^{2}}{b^{2}+T^{2}}\,,
\end{eqnarray}
\end{subequations}
which are both finite at $T=0$ for nonvanishing $b$
and a finite value of $\rho_{0}$.
Further details on this particular nonsingular-bouncing-cosmology
scenario can be found
in Refs.~\cite{Klinkhamer2019,KlinkhamerWang2019}.

For the numerical calculations with a dynamic scalar field
(to be introduced
in Sec.~\ref{subsec:Reduced-field-equations-with-dynamic-scalar-field}),
we will use the auxiliary
coordinate $\tau \in (-\infty,\,-b]  \, \cup \, [b,\,\infty)$
instead of $T \in \mathbb{R}$.
These two coordinates are related as follows:%
\begin{equation}
\label{eq:mod-FLRW-tau-of-T-def}
\hspace*{-0mm}
\tau(T)=
\begin{cases}
 + \sqrt{b^{2}+T^{2}}\,,    & \;\;\text{for}\;\; T \geq 0\,,
 \\[2mm]
 - \sqrt{b^{2}+T^{2}}\,,    & \;\;\text{for}\;\; T \leq 0\,,
\end{cases}
\end{equation}
where $\tau=-b$ and $\tau=b$
correspond to a single point ($T=0$) on the cosmic time axis
(see Ref.~\cite{Klinkhamer2019} for further discussion).

We remark that
the coordinate transformation from $T$ to $\tau$
is not a diffeomorphism (an invertible $C^{\infty}$ function):
the function \eqref{eq:mod-FLRW-tau-of-T-def}
is discontinuous between $T = 0^{-}$ and $T = 0^{+}$,
as is the (suitably defined) second derivative. This
implies that the metric \eqref{eq:mod-FLRW-ds2}
and the metric in terms of the $\tau$ coordinate
\eqref{eq:mod-FLRW-tau-of-T-def} give rise to different
differential structures of the respective spacetime manifolds
(see Refs.~\cite{Klinkhamer2019,KlinkhamerSorba2014,Guenther2017}
for further details).
Still, we can use the auxiliary $\tau$ coordinate
(with appropriate boundary conditions at $\tau = \pm b$)
to simplify the process of obtaining explicit solutions
of the field equations.

\subsection{Reduced field equations with a dynamic scalar field}
\label{subsec:Reduced-field-equations-with-dynamic-scalar-field}

We now consider a particular model for
a dynamic scalar field $\phi(x)$ propagating
over the spacetime manifold with  metric \eqref{eq:mod-FLRW-ds2}.
For the cosmological applications considered, the scalar field is
assumed to be spatially homogeneous and
to depend solely on the cosmic time coordinate, which is taken to be
the auxiliary coordinate $\tau$ from \eqref{eq:mod-FLRW-tau-of-T-def}.

The dynamic equations for the functions $\phi(\tau)$ and $a(\tau)$
are the Klein--Gordon equation, the second-order Friedmann equation,
and the first-order Friedmann equation,
\begin{subequations}\label{eq:ODEs}
\begin{eqnarray}
\label{eq:ODEs-ddot-phi}
\hspace*{-7.0mm}&&
\ddot{\phi}+3\,\left(\frac{\dot{a}}{a}\right)\,\dot{\phi}=
-\frac{\partial\,V_\text{eff}}{\partial\phi}\,\,,
\end{eqnarray}
\begin{eqnarray}
\label{eq:ODEs-ddot-a}
\hspace*{-7.0mm}&&
\frac{\ddot{a}}{a}=-\frac{8\pi G_N}{3}\,
\left(\dot{\phi}^{2} - V_\text{eff}
-\frac{1}{2}\,\frac{a}{\dot{a}}\,
\left[
\frac{d V_\text{eff}}{d \tau}\,
-\frac{\partial V_\text{eff}}{\partial \phi}\,\dot{\phi} \right]
\right)\,,
\end{eqnarray}
\begin{eqnarray}
\label{eq:ODEs-Friedmann}
\hspace*{-7.0mm}&&
\left(\frac{\dot{a}}{a}\right)^{2}=\frac{8\pi G_N}{3}\,
\left(\frac12\,\dot{\phi}^{2} + V_\text{eff} \right)\,,
\end{eqnarray}
\begin{eqnarray}
\label{eq:ODEs-Veff}
\hspace*{-7.0mm}&&
V_\text{eff}  =
\Big(1-\exp\left[\widehat{\alpha}^{6}-a^{6}\right]\Big)^{2}\,
\left(\frac{\text{sgn}\left[\dot{a}/a\right]+1}{2}\right)
\frac{1}{2}\,m^{2}\,\phi^{2} \,,
\\[2mm]
\label{eq:ODEs-phi-bcs}
\hspace*{-7.0mm}&&
\phi(b) = \phi(-b)\,,
\end{eqnarray}
\begin{eqnarray}
\label{eq:ODEs-a-bcs}
\hspace*{-7.0mm}&&
a^{2}(b) = a^{2}(-b) \equiv \widehat{\alpha}^{2}\,,
\end{eqnarray}
\end{subequations}
where the overdot stands for differentiation with respect to $\tau$
and the sign function is defined by
\begin{eqnarray}
\label{eq:sgn-def}
\hspace*{-0mm}
\text{sgn}(x)&\hspace{-2mm}=\hspace{-2mm}&
\begin{cases}
 x/\sqrt{x^{2}}\,,    & \;\;\text{for}\;\; x\ne 0\,,
 \\[2mm]
 0\,,    & \;\;\text{for}\;\; x=0\,.
\end{cases}
\end{eqnarray}
The boundary conditions \eqref{eq:ODEs-phi-bcs} and
\eqref{eq:ODEs-a-bcs} are supplemented with
boundary conditions on the derivatives $\dot{\phi}$ and $\dot{a}$
at $\tau=\pm b$, in order to have
well-defined functions $\phi(T)$ and $a(T)$ at $T=0$
(further details will be given
in Sec.~\ref{subsec:Numerical-analytic-results-model-phi}).

The effective potential \eqref{eq:ODEs-Veff} consists of the
standard quadratic term multiplied by two pre-factors with large brackets.
The second pre-factor in \eqref{eq:ODEs-Veff}
gives a vanishing potential in the contracting pre-bounce
phase and a nonvanishing potential in the expanding post-bounce
phase, while the first pre-factor makes for a smooth start
at $\tau=b$ of the nonvanishing potential in the post-bounce
phase ($\tau > b$).

The ordinary differential equations (ODEs) from
\eqref{eq:ODEs} are con\-sis\-tent,
as can be checked by calculating the $\tau$ derivative of the
first-order Friedmann equation \eqref{eq:ODEs-Friedmann} and
eliminating the obtained $\ddot{\phi}$ term
by use of the Klein--Gordon equation \eqref{eq:ODEs-ddot-phi}.
The resulting equation is
precisely the second-order Friedmann equation \eqref{eq:ODEs-ddot-a}
with the extra term on the right-hand side.
This extra term reads,
for the explicit choice  \eqref{eq:ODEs-Veff},
\begin{eqnarray}
\hspace*{-7mm}&&
\frac{8\pi G_N}{6}\,\frac{a}{\dot{a}}\,
\left[\dot{a}\;\frac{d}{d a}\,
\Big(1-\exp\left[\widehat{\alpha}^{6}-a^{6}\right]\Big)^{2}\right]\;
\left(\frac{\text{sgn}\left[\dot{a}/a\right]+1}{2}\right)\,
\nonumber\\[1mm]\hspace*{-7mm}&&
\times
\frac{1}{2}\,m^{2}\,\phi^{2}\,.
\end{eqnarray}
As mentioned before, 
further remarks on the effective potential \eqref{eq:ODEs-Veff}
appear in Sec.~\ref{sec:Discussion}.

For later use, we introduce the definitions
\begin{subequations}\label{eq:rho-phi-def-and-P-phi-def}
\begin{eqnarray}
\rho_{\phi} &\hspace{-2mm}\equiv\hspace{-2mm}& \frac12\,\dot{\phi}^{2} + \frac12\,m^{2}\,\phi^{2}\,,
\\[2mm]
P_{\phi} &\hspace{-2mm}\equiv\hspace{-2mm}& \frac12\,\dot{\phi}^{2} - \frac12\,m^{2}\,\phi^{2}\,,
\\[2mm]
w_{\phi} &\hspace{-2mm}\equiv\hspace{-2mm}& P_{\phi}/\rho_{\phi}\,,
\end{eqnarray}
\end{subequations}
which are primarily relevant in the post-bounce phase.
The two Friedmann equations \eqref{eq:ODEs-ddot-a}
and \eqref{eq:ODEs-Friedmann}
then become asymptotically ($\tau \gg b$):
\begin{subequations}\label{eq:ODEs-asymptotically}
\begin{eqnarray}
\label{eq:ODEs-ddot-a-asymptotically}
\left.\frac{\ddot{a}}{a}\,\right|^\text{(asymp.)}
&\hspace{-2mm}\sim\hspace{-2mm}&
-\frac{8\pi G_N}{3}\,
\left[ \frac{1}{2}\,\rho_{\phi} + \frac{3}{2}\,P_{\phi}\right]\,,
\\[2mm]
\label{eq:ODEs-Friedmann-asymptotically}
\left.\left(\frac{\dot{a}}{a}\right)^{2}\,\right|^\text{(asymp.)}
&\hspace{-2mm}\sim\hspace{-2mm}&
+\frac{8\pi G_N}{3}\;\rho_{\phi}\,,
\end{eqnarray}
\end{subequations}
which shows that the quantities $\rho_{\phi}$ and $P_{\phi}$
from \eqref{eq:rho-phi-def-and-P-phi-def} can be interpreted
as the energy density and the pressure of the
asymptotic homogeneous $\phi$ field~\cite{Mukhanov2005}.

Henceforth, we use reduced-Planckian units
and take explicit values for $b$ and $m$ in these units,
\begin{subequations}\label{eq:reduced-Planckian-units-AND-b-m-choices}
\begin{eqnarray}
\label{eq:reduced-Planckian-units}
\hspace*{-7mm}&&
8\pi G_N =c=\hbar=1\,,
\\[2mm]
\hspace*{-7mm}&&
b=1/m=1\,,
\end{eqnarray}
\end{subequations}
where $b$ and $1/m$ correspond, respectively, to the
length scale entering the metric \eqref{eq:mod-FLRW-ds2}
and the Compton wavelength of the $\phi$ scalar
in the post-bounce phase.

\begin{figure*}[t!]
\vspace*{-0mm}
\begin{center}
\includegraphics[width=0.95\textwidth]
{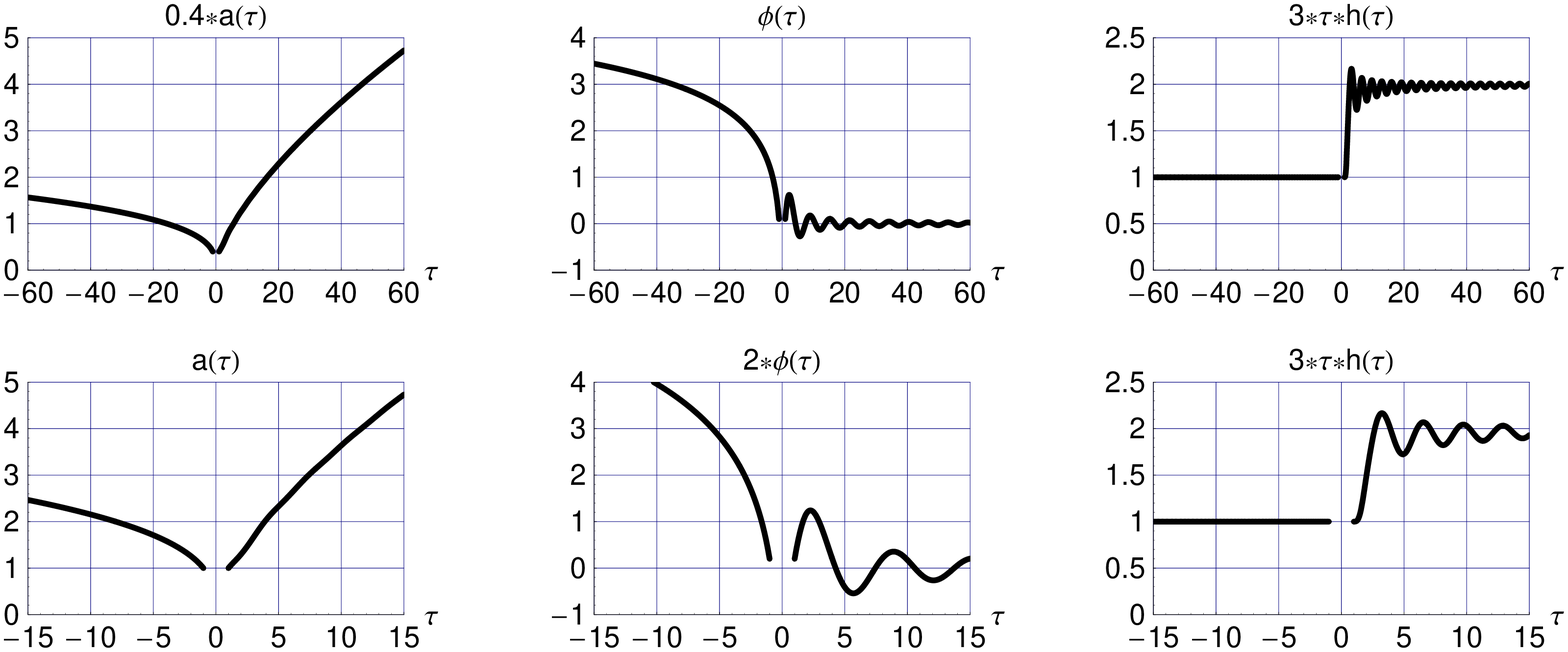}
\end{center}
\vspace*{-0mm}
\caption{Numerical solution of the second-order ODEs \eqref{eq:ODEs-ddot-phi}
and \eqref{eq:ODEs-ddot-a}, with boundary conditions satisfying the
first-order ODE \eqref{eq:ODEs-Friedmann}
for effective potential \eqref{eq:ODEs-Veff}.
The model parameters are $b=1/m=1$.
Shown are the dynamic variables $a(\tau)$ and
$\phi(\tau)$ for $|\tau| \geq b $, together with the corresponding
Hubble parameter $h(\tau) \equiv [d a(\tau)/d \tau]/a(\tau)$.
With $b=1$,  the boundary conditions are:
$a(1)=a(-1)=1$, $\dot{a}(1)=-\dot{a}(-1)=1/3$,
$\phi(1)=\phi(-1)=1/10$, and $\dot{\phi}(1)=-\dot{\phi}(-1)=
\sqrt{2/3} \approx 0.816497$.
}
\label{fig:fig1}
\vspace*{0mm}
%
\vspace*{0mm}
\begin{center}
\includegraphics[width=0.95\textwidth]
{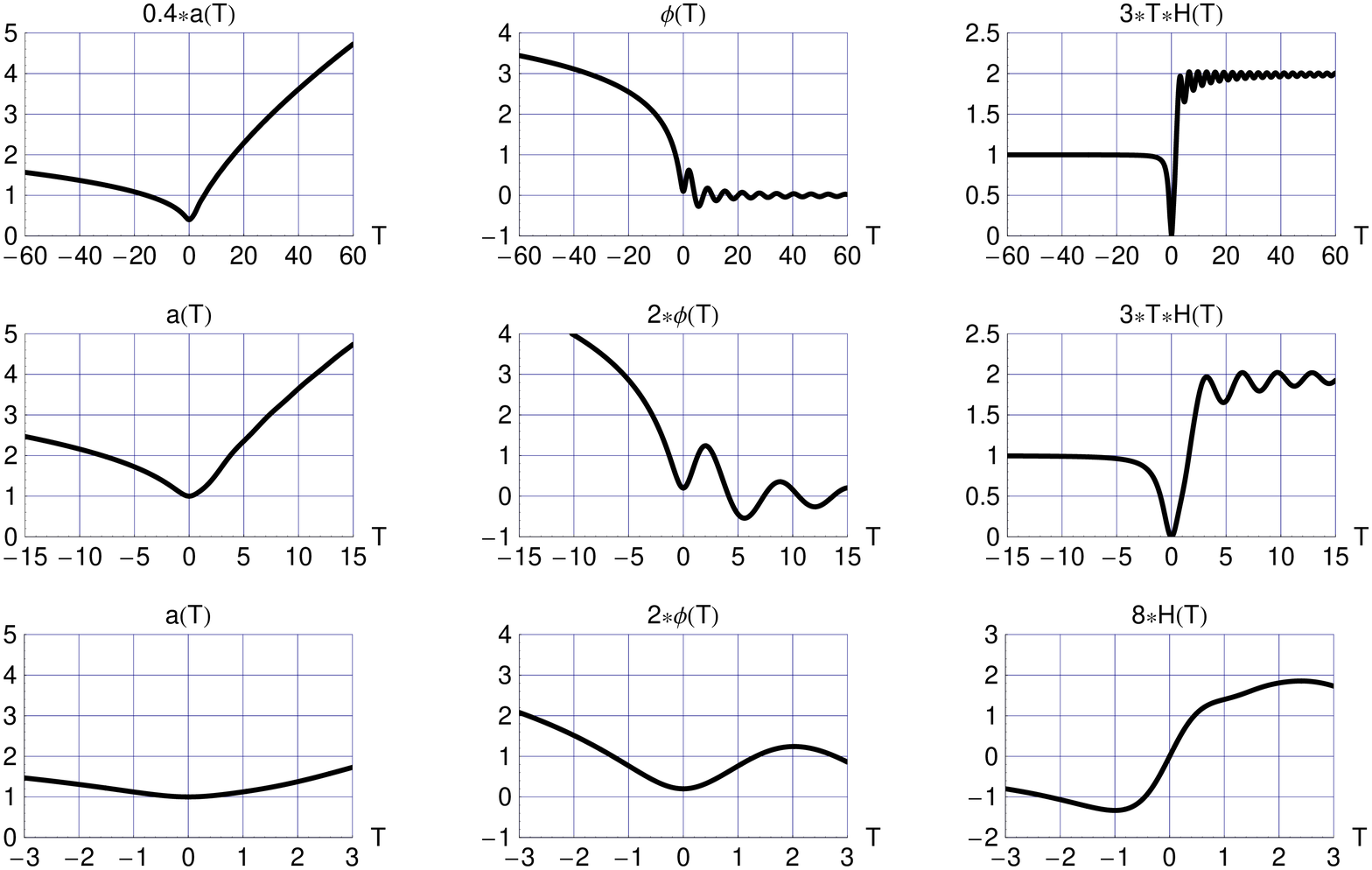}
\end{center}
\vspace*{-0mm}
\caption{Numerical solutions $a(\tau)$ and $\phi(\tau)$ from Fig.~\ref{fig:fig1}
plotted with respect to $T$ as defined by \eqref{eq:mod-FLRW-tau-of-T-def},
together with the corresponding
Hubble parameter $H(T) \equiv [d a(T)/d T]/a(T)$.
}
\label{fig:fig2}
\end{figure*}
\begin{figure*}[t]
\vspace*{-0mm}
\begin{center}
\includegraphics[width=0.6333333\textwidth]
{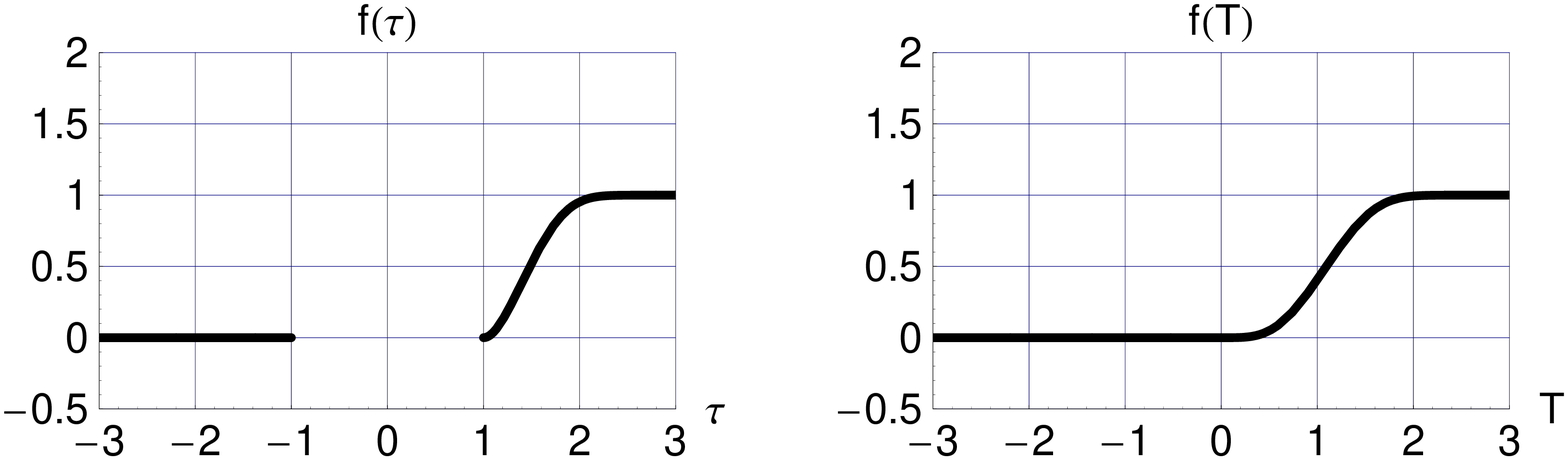}
\end{center}
\vspace*{-0mm}
\caption{On the left, prefactor $f(\tau)\equiv
\big(1-\exp\big[1-a^{6}(\tau)\big]\big)^{2}\,
\big(\text{sgn}\left[h(\tau)\right]+1\big)\big/2$
entering the effective potential \eqref{eq:ODEs-Veff}
for the numerical solution displayed in Fig.~\ref{fig:fig1}
and, on the right, the corresponding prefactor
\mbox{$f(T)\equiv
\big(1-\exp\big[1-a^{6}(T)\big]\big)^{2}\,
\big(\text{sgn}\left[H(T)\right]+1\big)\big/2$}
for the numerical solution displayed in Fig.~\ref{fig:fig2}.
}
\label{fig:fig3}
\vspace*{2mm}
\vspace*{-0mm}
\begin{center}
\includegraphics[width=0.95\textwidth]
{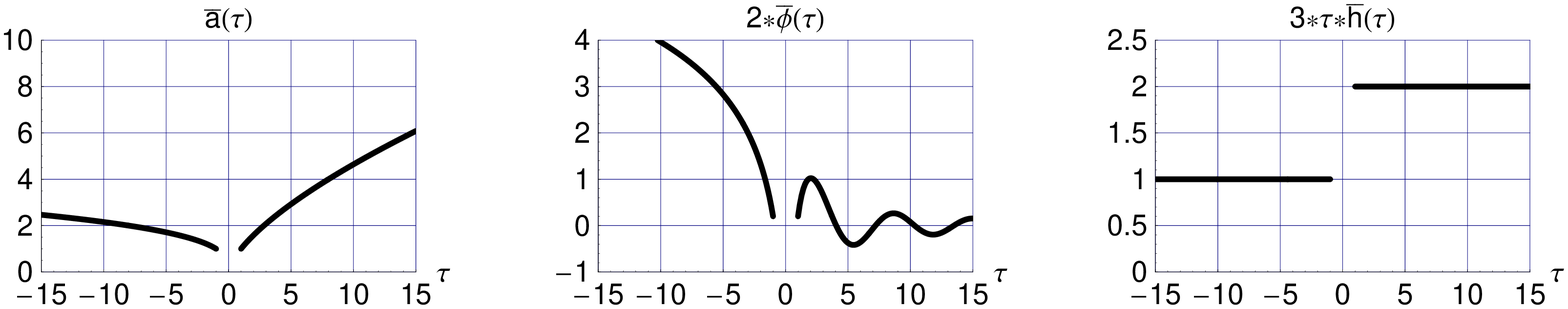}
\end{center}
\vspace*{-0mm}
\caption{Analytic expressions $\overline{a}(\tau)$, $\overline{\phi}(\tau)$, 
and $\overline{h}(\tau)$, as given by \eqref{eq:pre-bounce-analytic-sol}
and \eqref{eq:phi-h-asymp-sol} for $b=1/m=1$ with constants $c_{1}=1/10$
and $c_{2}=\arccos\big(-\sqrt{3}/20\big)-1+\pi\approx 3.7991$.
}
\label{fig:fig4}
\vspace*{-0mm}
\end{figure*}

\subsection{Numerical and analytic results}
\label{subsec:Numerical-analytic-results-model-phi}

We solve the ODEs \eqref{eq:ODEs} numerically.
Specifically, we solve the two
second-order equations \eqref{eq:ODEs-ddot-phi}
and \eqref{eq:ODEs-ddot-a},
with boundary conditions satisfying the
first-order Friedman equation \eqref{eq:ODEs-Friedmann}.

Numerical results are shown in Fig.~\ref{fig:fig1}
with the following boundary conditions at \mbox{$\tau=\pm b$\,:}
$a(b)=a(-b)$, $\dot{a}(b)=-\dot{a}(-b)$,
$\phi(b)=\phi(-b)$,
and $\dot{\phi}(b)=-\dot{\phi}(-b)$,
where the $\dot{\phi}^{2}$ value at $\tau=\pm b$ follows from \eqref{eq:ODEs-Friedmann}.
The top-left panel of Fig.~\ref{fig:fig1}, in particular,
makes clear that the bounce is time-asymmetric.

Two technical remarks are in order.
The first remark is that the ODEs are solved forward in cosmic time $\tau$
with the $\tau=b$ boundary conditions
and backward in cosmic time $\tau$ with the $\tau=-b$ boundary conditions.
The second remark is that we prefer to
work with the ODEs \eqref{eq:ODEs-ddot-phi}, \eqref{eq:ODEs-ddot-a},
and \eqref{eq:ODEs-Friedmann} in terms of
the auxiliary time coordinate $\tau$,
rather than the corresponding ODEs
in terms of the original time coordinate $T$.
The reason is that the $\tau$-ODEs are nonsingular equations,
whereas the \mbox{$T$-ODEs} are singular equations
\big[having, for example, a term
\mbox{$(1+b^{2}/T^{2})\,d^{2}\phi(T)/dT^{2}$}
in the Klein--Gordon equation\big].

As mentioned in the previous paragraph,
the results of Fig.~\ref{fig:fig1} are obtained with
the auxiliary cosmic time \mbox{coordinate $\tau$},
but the physically relevant
cosmic time coordinate is $T$ from \eqref{eq:mod-FLRW}.
Using \eqref{eq:mod-FLRW-tau-of-T-def},
the results from Fig.~\ref{fig:fig1} are re-plotted
with respect to  $T$ in Fig.~\ref{fig:fig2}:
the top row shows the asymptotic post-bounce behavior for $T \gg b$,
the middle row the unset of oscillatory behavior of the
scalar field $\phi(T)$ for $T > 0$,
and the bottom row the smoothness at $T=0$
(for further discussion on the smoothness,
see Ref.~\cite{Klinkhamer2019} and references therein).

The top-right panel in Fig.~\ref{fig:fig2} has,
in the asymptotic pre-bounce phase ($T \ll -b$),
a Hubble parameter \mbox{$H(T) \sim (1/3)\,T^{-1}$}
corresponding to the scale factor $a(T) \propto (T^{2})^{1/6}$
and, in the asymptotic post-bounce phase ($T \gg b$),
a Hubble parameter $H(T) \sim (2/3)\,T^{-1}$
corresponding to the scale factor $a(T) \propto T^{2/3}$.
This behavior results from having
effective equation-of-state parameters
$w_\text{eff,\,pre-bounce} \sim 1$
and $w_\text{eff,\,post-bounce} \sim 0$.
\mbox{Figure~\ref{fig:fig3}} shows, for the obtained numerical solution,
the pre\-factors entering the effective potential \eqref{eq:ODEs-Veff}.

We also have some analytic results.
In the pre-bounce phase ($\tau \leq -b < 0$),
we obtain the following analytic solution:%
\begin{subequations}\label{eq:pre-bounce-analytic-sol}
\begin{eqnarray}
\overline{\phi}_\text{pre-bounce}(\tau) &\hspace{-2mm}=\hspace{-2mm}&
\sqrt{1/6}\, \ln(\tau^{2}/b^{2}) + c_{1}\,,
\\[2mm]
\overline{h}_\text{pre-bounce}(\tau) &\hspace{-2mm}=\hspace{-2mm}& (1/3)\,\tau^{-1}\,,
\\[2mm]
\overline{a}_\text{pre-bounce}(\tau) &\hspace{-2mm}=\hspace{-2mm}& (\tau^{2}/b^{2})^{1/6}\,,
\end{eqnarray}
\end{subequations}
with an appropriate constant $c_{1}$ to match the
numerical value of $\phi(-1)$ from Fig.~\ref{fig:fig1}.
Notice that if the effective potential \eqref{eq:ODEs-Veff}
were absent (for example, from having $m=0$), the
nonsingular bouncing cosmology would be time-symmetric
with the behavior \eqref{eq:pre-bounce-analytic-sol}
over the whole cosmic time axis,
$\tau \in (-\infty,\,-b]  \, \cup \, [b,\,\infty)$.

Moreover, it is possible to obtain analytic results
for the asymptotic post-bounce behavior of the dimensionless variables
$\phi(\tau)$ and $h(\tau)$. \mbox{We consider $\tau \gg b$
with $V_\text{eff}  \sim (1/2)\,m^{2}\,\phi^{2}$.}
Making the \textit{Ansatz}
\begin{equation}
\label{eq:asymp-phi-Ansatz}
\phi(\tau)=\xi(\tau)\,\tau^{-1}
\end{equation}
and considering the resulting
$O(\tau^{-1})$ and $O(\tau^{-2})$ terms of the ODEs
\eqref{eq:ODEs-ddot-phi} and \eqref{eq:ODEs-Friedmann}, we obtain
the following asymptotic solution (for $\tau \gg b$):
\begin{subequations}\label{eq:phi-h-asymp-sol}
\begin{eqnarray}
\hspace*{-0mm}
\overline{\phi}_\text{post-bounce}(\tau)\,\Big|^\text{(asymp.)}
&\hspace{-2mm}\sim\hspace{-2mm}&
\frac{2}{\sqrt{3}}\;
\frac{\cos\big(m\tau+c_{2}\big)}{m\tau}\,,
\\[2mm]
\hspace*{-0mm}
\overline{h}_\text{post-bounce}(\tau)\,\Big|^\text{(asymp.)}
&\hspace{-2mm}\sim\hspace{-2mm}&
(2/3)\,\tau^{-1}\,,
\\[2mm]
\hspace*{-0mm}
\overline{a}_\text{post-bounce}(\tau)\,\Big|^\text{(asymp.)}
&\hspace{-2mm}\sim\hspace{-2mm}&
(\tau/b)^{2/3}\,,
\end{eqnarray}
\end{subequations}
with an appropriate constant $c_{2}$ to match the
numerical value of $\phi(1)$ from Fig.~\ref{fig:fig1}.
Higher-order terms are given by
Eqs.~(5.45) and (5.46) in Ref.~\cite{Mukhanov2005}.
In any case, we see that the appropriate \textit{Ansatz}
\eqref{eq:asymp-phi-Ansatz} inserted in the original ODEs
\eqref{eq:ODEs-ddot-phi} and \eqref{eq:ODEs-Friedmann}
already gives the leading terms of the asymptotic solution
for $\tau \gg b$.

Figure~\ref{fig:fig4} shows
the analytic behavior \eqref{eq:pre-bounce-analytic-sol}
and \eqref{eq:phi-h-asymp-sol} over the whole $\tau$ range,
even though the post-bounce results are only valid asymptotically.
The analytic results from Fig.~\ref{fig:fig4} may be compared with the
numerical results from the bottom-row panels in Fig.~\ref{fig:fig1}.

\begin{figure*}[t]
\vspace*{-2mm}
\begin{center}
\includegraphics[width=0.95\textwidth]
{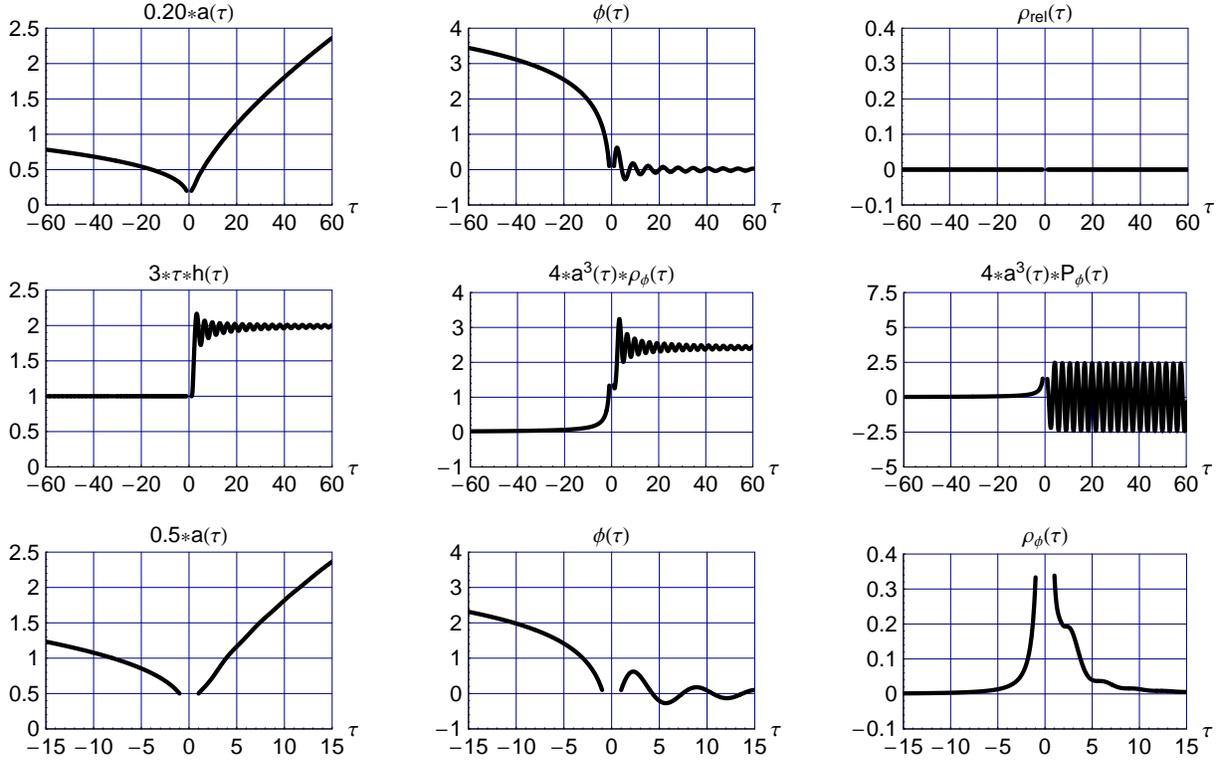}
\end{center}
\vspace*{-0mm}
\caption{Numerical solution of the ODEs
\eqref{eq:ODEs-ddot-phi-model-phi-rhorel}, \eqref{eq:ODEs-dot-rho-rel-model-phi-rhorel},
and \eqref{eq:ODEs-ddot-a-model-phi-rhorel},
with boundary conditions satisfying \eqref{eq:ODEs-Friedmann-model-phi-rhorel}
for effective potential \eqref{eq:ODEs-Veff-model-phi-rhorel}.
The model length scales are $b=1$ and $1/m=1$,
and the model decay constant vanishes, $\gamma=0$.
The top row shows
the dynamic variables $a(\tau)$, $\phi(\tau)$, and
$\rho_\text{rel}(\tau)$ for $|\tau| \geq b$.
The middle row shows the asymptotic post-bounce
behavior of certain derived quantities,
the Hubble parameter $h(\tau) \equiv [d a(\tau)/d \tau]/a(\tau)$
and the energy density $\rho_{\phi}(\tau)$ and pressure $P_{\phi}(\tau)$
as defined by \eqref{eq:rho-phi-def-and-P-phi-def}.
The bottom row shows the behavior near the spacetime defect
at $\tau=\pm b$ with the post-bounce onset of $\phi$
oscillations (solely dampened by the expansion of the
model universe). With $b=1$,  the boundary conditions are:
$a(1)=a(-1)=1$, $\dot{a}(1)=-\dot{a}(-1)=1/3$,
$\phi(1)=\phi(-1)=1/10$,
$\dot{\phi}(1)=-\dot{\phi}(-1)=\sqrt{2/3} \approx 0.816497$,
and $\rho_\text{rel}(1)=\rho_\text{rel}(-1)=0$.
}
\label{fig:fig5}
\vspace*{-0mm}
\end{figure*}

\section{Model with a massive scalar and relativistic matter}
\label{sec:Model-with-additional-relativistic-matter}

\setcounter{equation}{13} 

If the massive scalar field $\phi(x)$
of Sec.~\ref{sec:Model-with-massive-scalar}
is coupled to other fields, then the
oscillatory behavior of the post-bounce scalar field $\phi(x)$
gives rise to particle creation and reheating~\cite{Kofman-etal1997},
with $w_\text{eff}\sim 1/3$ from  massless or ultrarelativistic created particles.
In this section, we present a simplified calculation
for the decay of an oscillating massive scalar field
propagating over the spacetime manifold with
topology $\mathbb{R}^4$ and metric
\eqref{eq:mod-FLRW-ds2}.
An alternative calculation for a massless scalar with
quartic self-interactions is given in
\ref{app:Model with massless scalar and quartic self-interactions}.

All further calculations in this paper will be
performed solely with the auxiliary coordinate $\tau$
from \eqref{eq:mod-FLRW-tau-of-T-def}.
Still, the behavior of the cosmic scale factor $a(T)$,
the Hubble parameter $H(T)$, and matter fields such as $\phi(T)$
will be smooth at the defect surface $T=0$, as shown by the
bottom-row panels in Fig.~\ref{fig:fig2}.

In Sec.~\ref{subsec:Reduced-field-equations-model-phi-rhorel},
we consider the reduced field equations from a particular model
that incorporates the
energy exchange between the scalar field and
a relativistic matter component.
In Sec.~\ref{subsec:Numerical-results-model-phi-rhorel}, we obtain
the numerical solution of these reduced field equations.

\subsection{Reduced field equations}
\label{subsec:Reduced-field-equations-model-phi-rhorel}

The rapid oscillations of the post-bounce scalar field $\phi$
in Fig.~\ref{fig:fig2}
are expected to decay rapidly~\cite{Kofman-etal1997},
as long as the scalar particle $\phi$ of mass $m>0$
is coupled to light particles
(for example, a scalar particle $\chi$ of mass $\mu\,m$
with $0 \leq \mu\ll 1$). A coupling  term
$g\, m\, \phi\, \chi^{2}$ in the Lagrange density gives
a tree-level decay rate $\Gamma = g^{2}\, m/ (8 \pi)$
for the process $\phi \to \chi\,\chi$.
In a cosmological context, there are other effects
which may increase the effective decay rate~\cite{Mukhanov2005},
but, here, we only intend to present a simplified (but consistent)
model.

Our particular model involves the bounce field $\phi(\tau)$
together with a relativistic matter component
which is described by a homogeneous perfect fluid
with energy density $\rho_\text{rel}(\tau)$
and pressure $P_\text{rel}(\tau)= (1/3)\,\rho_\text{rel}(\tau)$.
In fact, the model extends the discussion of
Sec. 5.5.1 in Ref.~\cite{Mukhanov2005} and aims to give a
consistent description of the post-bounce evolution.

\begin{figure*}[t]
\vspace*{-0mm}
\begin{center}
\includegraphics[width=0.95\textwidth]
{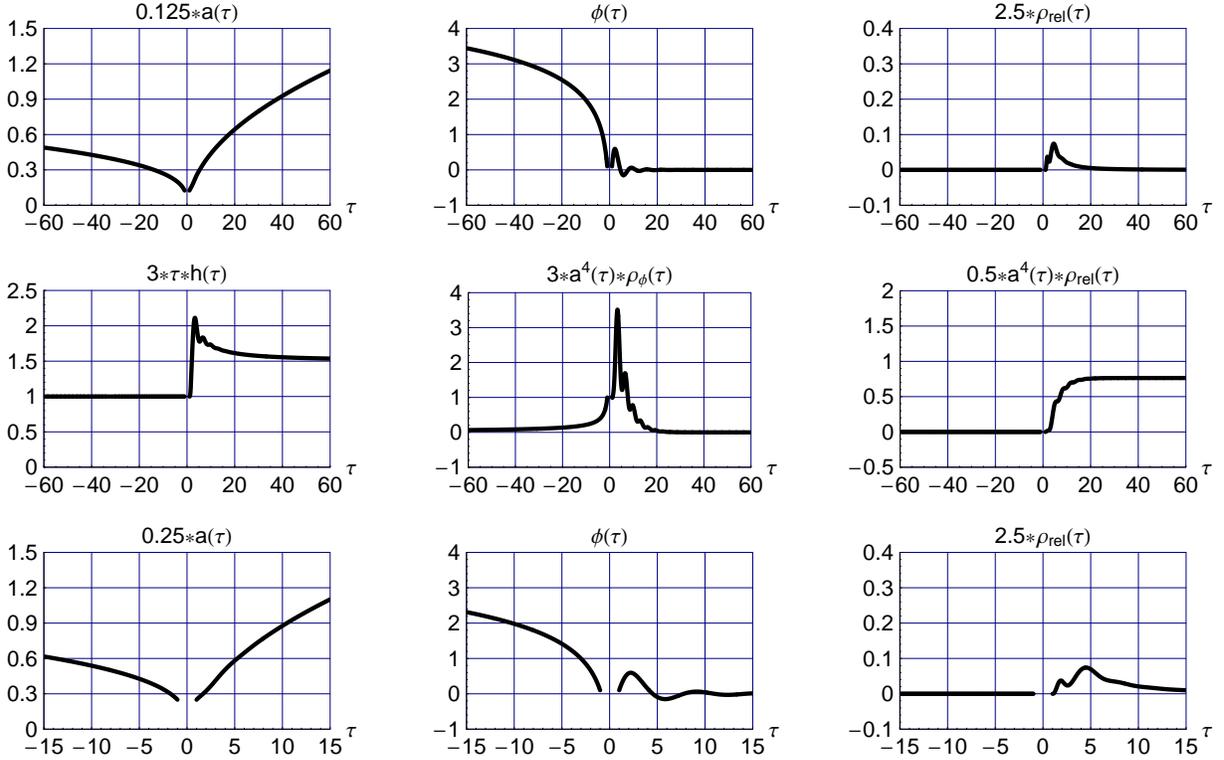}
\end{center}
\vspace*{-0mm}
\caption{Same as Fig.~\ref{fig:fig5}, but now with a nonvanishing
decay constant, $\gamma=1/10$.
}
\label{fig:fig6}
\vspace*{-0mm}
\end{figure*}
\begin{figure*}[t]
\vspace*{-0mm}
\begin{center}
\includegraphics[width=0.95\textwidth]
{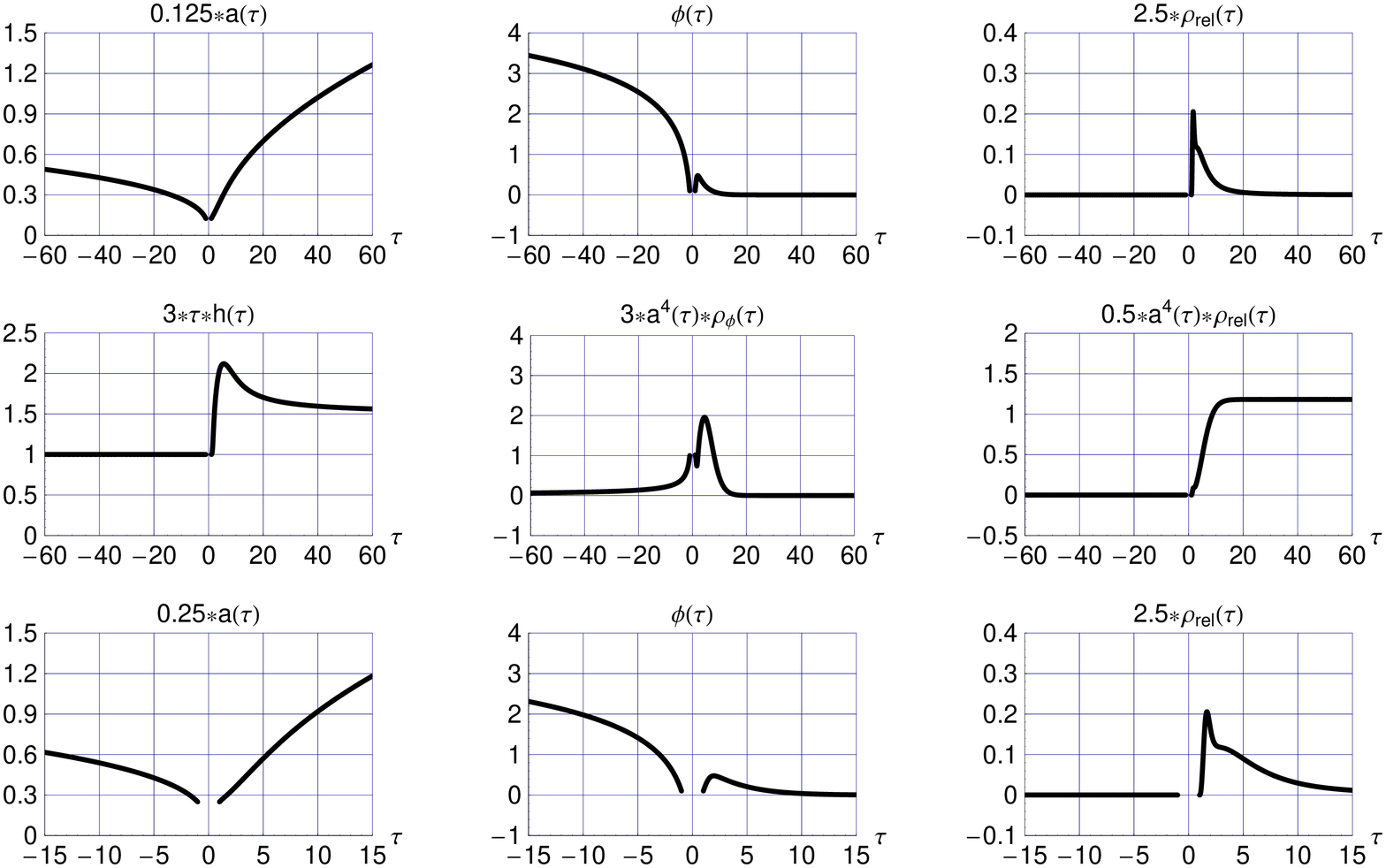}
\end{center}
\vspace*{-0mm}
\caption{Same as Fig.~\ref{fig:fig6}, but now with a larger
decay constant, $\gamma=1$.
}
\label{fig:fig7}
\vspace*{-0mm}
\end{figure*}

In addition to the homogeneous
scalar field $\phi(\tau)$  responsible for the bounce,
we thus consider a homogeneous relativistic matter component with
a constant equation-of-state parameter,
\begin{equation}
\label{eq:w-rel-def-model-phi-rhorel}
w_\text{rel}(\tau)
\equiv
P_\text{rel}(\tau)/\rho_\text{rel}(\tau)
=1/3\,.
\end{equation}
We now proceed in three steps.
First, we modify the Klein--Gordon equation by the introduction
of a friction term~\cite{Mukhanov2005} with decay constant $\Gamma$,
\begin{equation}
\label{eq:modified-KG-equation-model-phi-rhorel}
\ddot{\phi}
+3\,\big( \dot{a}/a + \Gamma \big)\,\dot{\phi} +m^{2}\,\phi = 0\,,
\end{equation}
where the overdot stands for differentiation with respect to $\tau$.
Second, we obtain the corresponding evolution equation for $\rho_{\phi}$
from \eqref{eq:modified-KG-equation-model-phi-rhorel}
and the definitions \eqref{eq:rho-phi-def-and-P-phi-def}
for $\rho_{\phi}$  and $P_{\phi}$,
\begin{equation}
\label{eq:dot-rho-phi-eq-model-phi-rhorel}
\dot{\rho}_{\phi} + 3\, (\dot{a}/a)\, \big(\rho_{\phi}+P_{\phi}\big)
= - 3\,\Gamma\, \big(\rho_{\phi}+P_{\phi}\big)\,.
\end{equation}
Third, energy conservation then gives the evolution
equation for $\rho_\text{rel}$,
\begin{equation}
\label{eq:dot-rho-rel-eq-model-phi-rhorel}
\dot{\rho}_\text{rel} + 4\, \big(\dot{a}/a\big)\, \rho_\text{rel}
= + 3\,\Gamma\, \big(\rho_{\phi}+P_{\phi}\big)\,,
\end{equation}
where the right-hand side can also be written
as $3\,\Gamma\,\dot{\phi}^{2}$.
The terms on the right-hand sides
of \eqref{eq:dot-rho-phi-eq-model-phi-rhorel}
and \eqref{eq:dot-rho-rel-eq-model-phi-rhorel}
describe the energy exchange between the scalar-field matter component
and a relativistic matter component characterized by
$\rho_\text{rel}$ and $P_\text{rel}=(1/3)\,\rho_\text{rel}$.

Replacing the decay constant $\Gamma$
in \eqref{eq:modified-KG-equation-model-phi-rhorel}
and \eqref{eq:dot-rho-rel-eq-model-phi-rhorel}
by a time-dependent quantity $\Gamma_\text{eff}$,
the following consistent ODES are obtained:
\begin{subequations}\label{eq:ODEs-model-phi-rhorel}
\begin{eqnarray}
\label{eq:ODEs-ddot-phi-model-phi-rhorel}
\hspace*{-7.00mm}&&
\ddot{\phi}+3\,\left(\frac{\dot{a}}{a}+\Gamma_\text{eff}\right)\,\dot{\phi}
=
-\frac{\partial\,V_\text{eff}}{\partial\phi}\,\,,
\\[2mm]
\label{eq:ODEs-dot-rho-rel-model-phi-rhorel}
\hspace*{-7.00mm}&&
\dot{\rho}_\text{rel} + 4\, \big(\dot{a}/a\big)\, \rho_\text{rel}
=
3\,\Gamma_\text{eff}\; \dot{\phi}^{2}\,,
\\[2mm]
\label{eq:ODEs-ddot-a-model-phi-rhorel}
\hspace*{-7.00mm}&&
\frac{\ddot{a}}{a}=-\frac{8\pi G_N}{3}\,
\Bigg( \dot{\phi}^{2} - V_\text{eff}+\rho_\text{rel}
-\frac{1}{2}\,\frac{a}{\dot{a}}\,
\left[\frac{d V_\text{eff}}{d \tau}\,
-\frac{\partial V_\text{eff}}{\partial \phi}\,\dot{\phi} \right]
\Bigg)\,,
\nonumber\\[1mm]\hspace*{-7.00mm}&&
\\[2mm]
\label{eq:ODEs-Friedmann-model-phi-rhorel}
\hspace*{-7.00mm}&&
\left(\frac{\dot{a}}{a}\right)^{2}=\frac{8\pi G_N}{3}\,
\left(\frac12\,\dot{\phi}^{2} + V_\text{eff} + \rho_\text{rel}\right)\,,
\\[2mm]
\label{eq:ODEs-Veff-model-phi-rhorel}
\hspace*{-7.00mm}&&
V_\text{eff}
=
f\,\frac{1}{2}\,m^{2}\,\phi^{2} \,,
\\[2mm]
\label{eq:ODEs-Gammaeff-model-phi-rhorel}
\hspace*{-7.00mm}&&
\Gamma_\text{eff}
=
f\,\gamma\,m \,,
\\[2mm]
\label{eq:ODEs-f-def-model-phi-rhorel}
\hspace*{-7.00mm}&&
f(\tau) \equiv
\Big(1-\exp\left[\widehat{\alpha}^{6}-a^{6}(\tau)\right]\Big)^{2}\,
\left(\frac{\text{sgn}\left[\dot{a}(\tau)/a(\tau)\right]+1}{2}\right)\,,
\nonumber\\[1mm]\hspace*{-7.00mm}&&
\end{eqnarray}
\begin{eqnarray}
\label{eq:ODEs-phi-bcs-model-phi-rhorel}
\hspace*{-7.00mm}&&
\phi(b) = \phi(-b)\,,
\\[2mm]
\label{eq:ODEs-a-bcs-model-phi-rhorel}
\hspace*{-7.00mm}&&
a^{2}(b) = a^{2}(-b) \equiv \widehat{\alpha}^{2}\,,
\end{eqnarray}
\end{subequations}
with Newton's constant $G_{N}$ temporarily displayed and
a nonnegative coupling constant $\gamma$ in $\Gamma_\text{eff}$
(the $\phi$-scalar mass $m$ is taken to be positive).
Two technical remarks are in order.
First, the additional term $\rho_\text{rel}$
on the right-hand side of \eqref{eq:ODEs-ddot-a-model-phi-rhorel}
corresponds to the extra contribution
$(1/2)\,\rho_\text{rel}+(3/2)\,P_\text{rel}\,$,
just as in the ODE \eqref{eq:ODEs-ddot-a-asymptotically}
for the $\phi$ field.
Second, more or less the same numerical results are obtained
without the smoothing factor
$\big(1-\exp\big[\widehat{\alpha}^{6}-a^{6}\big]\big)^{2}$
in $\Gamma_\text{eff}$ from \eqref{eq:ODEs-Gammaeff-model-phi-rhorel}
and \eqref{eq:ODEs-f-def-model-phi-rhorel}.

\vspace*{-0mm}
\subsection{Numerical results}
\label{subsec:Numerical-results-model-phi-rhorel}
\vspace*{-0mm}

We have obtained numerical solutions of the ODEs
\eqref{eq:ODEs-model-phi-rhorel}
for nonvanishing decay coupling constant $\gamma$
and with the same boundary conditions as in Fig.~\ref{fig:fig1}.
For comparison, we first show in Fig.~\ref{fig:fig5}
(which partly reproduces Fig.~\ref{fig:fig1})
the numerical solution without $\phi$ decay ($\gamma=0$).
The $\phi$ oscillations in the post-bounce phase give
a vanishing average pressure $P_{\phi}$
and an average energy density $\rho_{\phi} \propto 1/a^3$,
as shown by the right and mid panels
of the middle row in Fig.~\ref{fig:fig5}.
The resulting post-bounce expansion has
$a(\tau) \propto \tau^{2/3}$ with
Hubble parameter $h(\tau) \sim (2/3)\,\tau^{-1}$,
as shown by the left panel of the middle row in Fig.~\ref{fig:fig5}.

With decay coupling constants $\gamma=1/10$ and $\gamma=1$,
Figs.~\ref{fig:fig6} and \ref{fig:fig7} show that
the post-bounce $\phi$ oscillations are rapidly damped
(with $\rho_{\phi}$ dropping to zero even faster than $1/a^4$)
and that the $\phi$-oscillation energy is completely transferred
to the relativistic component $\rho_\text{rel} \propto 1/a^4$.
The resulting post-bounce expansion has $a(\tau) \propto \tau^{1/2}$
with Hubble parameter \mbox{$h(\tau) \sim (1/2)\,\tau^{-1}$}.
Observe that the behavior of the post-bounce $\phi$-oscillations
in the bottom-mid panels of Figs.~\ref{fig:fig5}--\ref{fig:fig7}
ranges from underdamped to overdamped.

\section{Discussion}
\label{sec:Discussion}

The construction of the scalar-field model for the asymmetric nonsingular bounce
in Sec.~\ref{sec:Model-with-massive-scalar}
(and also in \ref{app:Model with massless scalar and quartic self-interactions})
is relatively straightforward, but there is a subtle point.

Indeed, the crucial property of the effective potential $V_\text{eff}(\phi)$
in \eqref{eq:ODEs-Veff} is its time-reversal-noninvariance:
the effective potential vanishes in the contracting phase
and is nonvanishing in the expanding phase.
A possible origin of such an effective potential may come
from a fundamental arrow-of-time, one example
being the anomalous time-reversal violation~\cite{Klinkhamer2002}
obtained from the chiral $SO(10)$ gauge theory containing the
three-family Standard Model defined over a manifold
with topology $\mathbb{R}\times T^3$
(see Ref.~\cite{Klinkhamer2000} for the original paper on the
CPT anomaly and Ref.~\cite{Klinkhamer2018} for a recent review).

The idea is that $V_\text{eff}(\phi)$ in \eqref{eq:ODEs-Veff}
originates from quantum effects, where the CPT
anomaly~\cite{Klinkhamer2002,Klinkhamer2000,Klinkhamer2018}
is responsible for a microscopic ``time-direction''
(giving the second prefactor with the sign function
in $V_\text{eff}$)
and where gravitational effects involving $H^{2}(T)$ are
responsible for the smooth turn-on
(giving  the first prefactor with the exponential function
in $V_\text{eff}$).
The outstanding task is to calculate such an effective potential
\mbox{\emph{ab initio}}.
Note that previous results for particle creation in
nonsingular bouncing cosmology appear to rely on time-reversal-noninvariant
boundary conditions; see, e.g., Ref.~\cite{Quintin-etal2014}.

Another issue is the physical interpretation of
the fundamental scalar field $\phi(x)$ used in
Secs.~\ref{sec:Model-with-massive-scalar}
and \ref{sec:Model-with-additional-relativistic-matter}.
Perhaps it is possible to interpret
this fundamental scalar field $\phi(x)$ [or rather an extra copy of it]
as the fluctuating component $\xi(x)$ of the
composite scalar field $q(x)=q_{0}+\xi(x)$ from the
so-called $q$-theory approach to the cosmological constant problem (see
Refs.~\cite{KV2008a,KV2016-Lambda-cancellation,KV2016-q-DM,%
KV2016-More-on-q-DM,KlinkhamerMistele2017}
and references therein).
The results of Sec.~\ref{sec:Model-with-additional-relativistic-matter}
then illustrate, for an assumed 
nonzero 
and positive value of $\gamma$,
how the oscillating homogeneous component $\xi(x)$ of the
$q$-field transfers its energy to relativistic particles in the
post-bounce phase (this energy-transfer process differs from the
one considered in Ref.~\cite{KV2016-Lambda-cancellation}).

\section*{Acknowledgments}
We thank the referee for constructive remarks.
The work of Z.L.W. is supported by the China Scholarship Council.

\begin{appendix}
\section{Model with a massless scalar and quartic\\ self-interactions}
\label{app:Model with massless scalar and quartic self-interactions}

\begin{figure*}[t]
\vspace*{-0mm}
\begin{center}
\includegraphics[width=0.95\textwidth]
{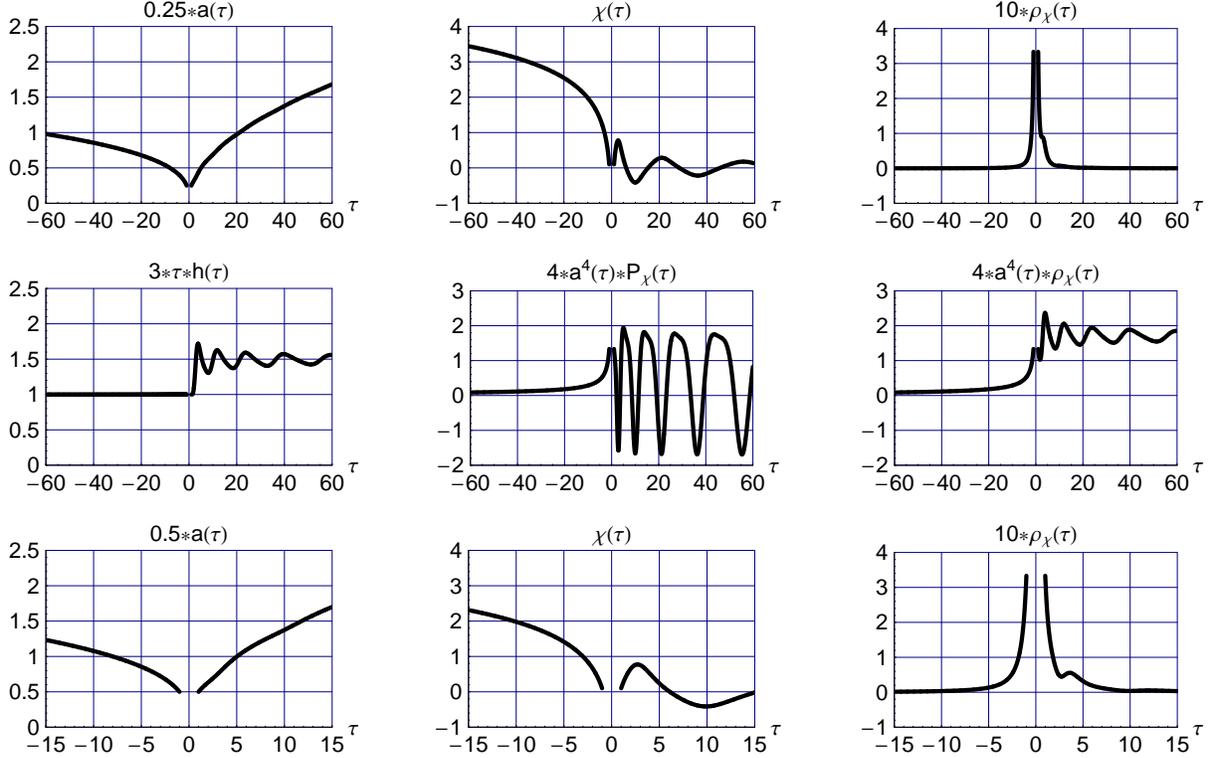}
\end{center}
\vspace*{-0mm}
\caption{Numerical solution of the ODEs \eqref{eq:ODEs-ddot-phi}
and \eqref{eq:ODEs-ddot-a} with $\phi(\tau)$ replaced by $\chi(\tau)$.
The boundary conditions satisfy \eqref{eq:ODEs-Friedmann}
in terms of $\chi(\tau)$,
for effective potential \eqref{eq:ODEs-Veff-appA}.
The model length scale is given by $b=1$
and the quartic coupling constant by $\lambda=1$.
The top row shows the dynamic variables $a(\tau)$ and $\chi(\tau)$
for $|\tau| \geq b$,
together with $\rho_{\chi}(\tau)$ from \eqref{eq:rho-chi-def-appA}.
The middle row shows the asymptotic post-bounce
behavior of certain derived quantities,
the Hubble parameter $h(\tau) \equiv [d a(\tau)/d \tau]/a(\tau)$
and the energy density $\rho_{\chi}(\tau)$ and
pressure $P_{\chi}(\tau)$ as defined by
\eqref{eq:rho-chi-def-and-P-chi-def-w-chi-def-appA}.
The bottom row shows the behavior near the spacetime defect
at $\tau=\pm b$ with the post-bounce onset of $\chi$
oscillations. With $b=1$,  the boundary conditions are:
$a(1)=a(-1)=1$, $\dot{a}(1)=-\dot{a}(-1)=1/3$,
$\chi(1)=\chi(-1)=1/10$, and
$\dot{\chi}(1)=-\dot{\chi}(-1)=\sqrt{2/3} \approx 0.816497$.
}
\label{fig:fig8}
\end{figure*}
\vspace*{0mm}
\begin{figure*}[t]
\vspace*{-0mm}
\begin{center}
\includegraphics[width=0.95\textwidth]
{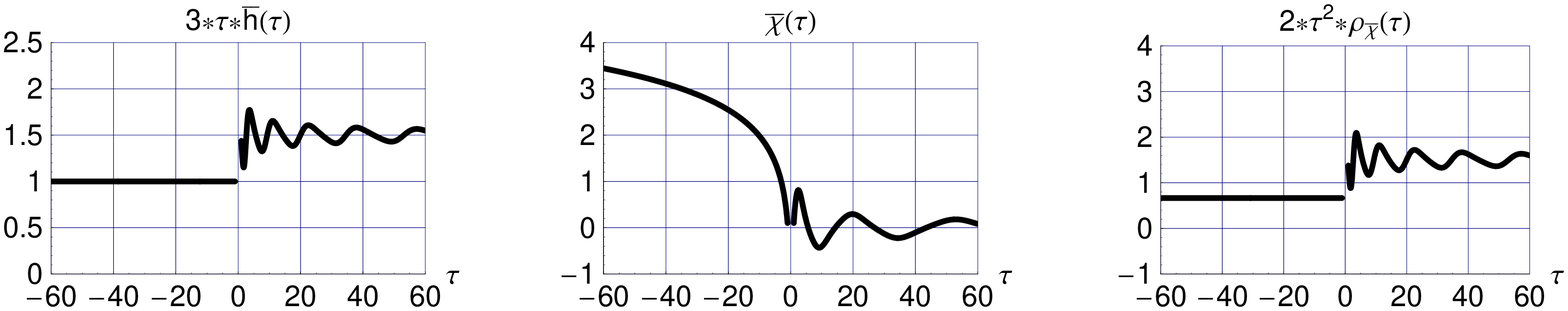}
\end{center}
\vspace*{-0mm}
\caption{Asymptotic post-bounce solution \eqref{eq:asymptotic-sols-appA}
with $\lambda=1$ and $c_{3} \approx 1.78523$.
Also shown is the analytic pre-bounce solution
\eqref{eq:pre-bounce-analytic-sol}
with $\overline{\phi}(\tau)$ replaced by $\overline{\chi}(\tau)$
and $\lambda=1$ and $c_{1} =1/10$.
}
\label{fig:fig9}
\vspace*{-0mm}
\end{figure*}

In Sec.~\ref{sec:Model-with-massive-scalar}, we considered
a massive free scalar $\phi$ in the post-bounce phase.
Here, we take a massless scalar $\chi$
with quartic self-interactions in the post-bounce phase.
The spacetime manifold has, again, topology $\mathbb{R}^4$
and metric \eqref{eq:mod-FLRW-ds2}.
The actual calculations use the auxiliary cosmic time
coordinate $\tau$ from \eqref{eq:mod-FLRW-tau-of-T-def} and the
overdot stands for differentiation with respect to $\tau$.

Specifically, we have the ODEs \eqref{eq:ODEs-ddot-phi},
\eqref{eq:ODEs-ddot-a}, and \eqref{eq:ODEs-Friedmann} with
$\phi(\tau)$ replaced by $\chi(\tau)$ and the following effective potential:
\begin{equation}
\label{eq:ODEs-Veff-appA}
V_\text{eff} =
\Big(1-\exp\left[\widehat{\alpha}^{6}-a^{6}\right]\Big)^{2}\,
\left(\frac{\text{sgn}\left[\dot{a}/a\right]+1}{2}\right)
\,\frac{\lambda}{4}\,\chi^{4} \,,
\end{equation}
with $\widehat{\alpha}^{2} \equiv a^{2}(b) = a^{2}(-b)$
and a positive coupling constant $\lambda$. Also define
the following quantities:
\begin{subequations}\label{eq:rho-chi-def-and-P-chi-def-w-chi-def-appA}
\begin{eqnarray}\label{eq:rho-chi-def-appA}
\rho_{\chi} &\hspace{-2mm}\equiv\hspace{-2mm}& \frac12\,\dot{\chi}^{2} + \frac{\lambda}{4}\,\chi^{4}\,,
\end{eqnarray}
\begin{eqnarray}
P_{\chi} &\hspace{-2mm}\equiv\hspace{-2mm}& \frac12\,\dot{\chi}^{2} - \frac{\lambda}{4}\,\chi^{4}\,,
\end{eqnarray}
\begin{eqnarray}
w_{\chi} &\hspace{-2mm}\equiv\hspace{-2mm}& P_{\chi}/\rho_{\chi}\,,
\end{eqnarray}
\end{subequations}
which are primarily relevant in the post-bounce phase.
Numerical results are presented in Fig.~\ref{fig:fig8}.
The left panel of the middle row in Fig.~\ref{fig:fig8}
shows that the post-bounce expansion $a \propto \tau^{1/2}$
for $\tau\gg b$ \big[with
$h \equiv \dot{a}/a \sim (1/2)\,\tau^{-1}$\big]
resembles the one of a standard radiation-dominated FLRW
universe ($w_\text{matter}=1/3$),
as noted already in Sec.~5.4.2. of Ref.~\cite{Mukhanov2005}.

The analytic pre-bounce solution is given by
\eqref{eq:pre-bounce-analytic-sol}
with $\overline{\phi}(\tau)$ replaced by $\overline{\chi}(\tau)$.
We will now get the asymptotic post-bounce
expressions for the dimensionless variables $\chi(\tau)$ and $h(\tau)$.
Just as for the case of the quadratic potential discussed
in the penultimate paragraph
of Sec.~\ref{subsec:Numerical-analytic-results-model-phi},
the crucial step is to make an appropriate \textit{Ansatz}
for $\chi(\tau)$.

Using reduced-Planckian units \eqref{eq:reduced-Planckian-units},
the relevant equations for $\chi(\tau)$ and $h(\tau)$ are
\begin{subequations}\label{eq:asymptotic-ODEs-appA}
\begin{eqnarray}
\label{eq:asymptotic-ODEs-ddot-chi-appA}
\hspace*{-7.000mm}&&
\ddot{\chi}+3\,h\,\dot{\chi}= -\lambda\,\chi^3\,,
\\[2mm]
\label{eq:asymptotic-ODEs-Friedmann-appA}
\hspace*{-7.000mm}&&
h^{2}=\frac{1}{3}\,
\left(\frac12\,\dot{\chi}^{2} + \frac{\lambda}{4}\,\chi^4 \right)\,.
\end{eqnarray}
\end{subequations}
The asymptotic solution $\chi_\text{asymp}(\tau)$ for $\tau \gg b > 0$
will be seen to fluctuate around zero and the asymptotic solution
$h_\text{asymp}(\tau)$ around $(1/2)\,\tau^{-1}$.

The procedure consists of two steps. First, we modify
the ODE \eqref{eq:asymptotic-ODEs-ddot-chi-appA}
by replacing $h(\tau)$ with $(1/2)\,\tau^{-1}$ and we get
\begin{equation}
\label{eq:MODIFIED-asymptotic-ODEs-ddot-chi-appA}
\left.\left(\ddot{\chi}+\frac{3}{2}\,\frac{1}{\tau}\,\dot{\chi}
+\lambda\,\chi^3\right)\,\right|^\text{(asymp.)}=0\,.
\end{equation}
Second, we make the following \textit{Ansatz}:
\begin{subequations}\label{eq:asymptotic-chi-Ansatz-rho-def-appA}
\begin{eqnarray}
\chi(\tau) &\hspace{-2mm}=\hspace{-2mm}& \eta(\rho)\;\frac{2}{\rho}\,,
\\[2mm]
\rho(\tau)  &\hspace{-2mm}\equiv\hspace{-2mm}& 2\,\sqrt{\tau}\,,
\end{eqnarray}
\end{subequations}
where $\eta(\rho)$ fluctuates around zero.
From \eqref{eq:MODIFIED-asymptotic-ODEs-ddot-chi-appA}
and \eqref{eq:asymptotic-chi-Ansatz-rho-def-appA}, we obtain
\begin{equation}
\label{eq:TMP-MODIFIED-asymptotic-ODEs-primeprime-eta-appA}
\frac{1}{\tau ^{3/2}}\,\Big(\eta ^{\prime \prime} + \lambda\, \eta ^3\Big) = 0\,,
\end{equation}
where the prime stands for differentiation with respect to $\rho$.
As the auxiliary cosmic time coordinate $\tau$ is nonvanishing for nonzero
defect scale $b$,
\eqref{eq:TMP-MODIFIED-asymptotic-ODEs-primeprime-eta-appA} reduces to
\begin{equation}
\label{eq:MODIFIED-asymptotic-ODEs-primeprime-eta-appA}
\eta ^{\prime \prime} + \lambda\, \eta ^3 = 0\,,
\end{equation}
which corresponds to a nonlinear second-order ODE
(in fact, a type of Bellman's equation,
$d^{2} y/dx^{2} = k\, x^l\, y^m$ for $k=-\lambda$, $l=0$, and $m=3$).

The solution of the ODE \eqref{eq:MODIFIED-asymptotic-ODEs-primeprime-eta-appA}
is given by a Jacobian elliptic function $\text{sn}(u\,|\,m)$,
in the notation of Ref.~\cite{AbramowitzStegun1965}.
Taking the two integration constants appropriately (see below),
the asymptotic ($\tau \gg b$) post-bounce solutions of $\chi(\tau)$ and $h(\tau)$
are given by
\begin{subequations}\label{eq:asymptotic-sols-appA}
\begin{eqnarray}
\label{eq:asymptotic-sol-chi-appA}
\overline{\chi}(\tau)\,\Big|^\text{(asymp.)}
&\hspace{-2mm}=\hspace{-2mm}&
\sqrt[4]{3/\lambda}\;\;
\text{sn}\Big(\sqrt[4]{12\,\lambda}\;\sqrt{\tau}+c_{3}\,\Big|\,-1\Big)
\;\;\frac{1}{\sqrt{\tau}}\,,
\nonumber\\[1mm]&&
\end{eqnarray}
\begin{eqnarray}
\label{eq:asymptotic-sol-h2-appA}
\overline{h}^{2}(\tau)\,\Big|^\text{(asymp.)}
&\hspace{-2mm}=\hspace{-2mm}&\frac{1}{3}\;
\left(\frac12\,\dot{\chi}_\text{asymp}^{2}(\tau)
+ \frac{\lambda}{4}\,\chi_\text{asymp}^4(\tau) \right)\,,
\end{eqnarray}
\end{subequations}
with a real constant $c_{3}$ that is, for the moment, set to zero.
The leading behavior of $h^{2}$ from \eqref{eq:asymptotic-sol-h2-appA},
for the $\chi$ solution \eqref{eq:asymptotic-sol-chi-appA}
with $c_{3}=0$,
equals $(1/4)\,\tau^{-2}$ upon use of the
identity $\text{cn}^{2}\,\text{dn}^{2} + \text{sn}^4=1$, which
holds~\cite{AbramowitzStegun1965} for parameter $m=-1$.
The asymptotic behavior $h^{2} \sim (1/4)\,\tau^{-2}$
explains \textit{a posteriori} the particular choice of one of the two
integration constants needed
to get \eqref{eq:asymptotic-sol-chi-appA} with $c_{3}=0$,
the other integration constant is chosen to get the simplest
possible argument of the $\text{sn}$ function
(an argument just proportional to $\sqrt{\tau}$).
Finally, we observe
from \eqref{eq:MODIFIED-asymptotic-ODEs-primeprime-eta-appA} that
the $\rho$ variable in a particular $\eta(\rho)$ solution
can be shifted by an arbitrary constant
and we determine an appropriate real constant $c_{3}$
in \eqref{eq:asymptotic-sol-chi-appA} to match the
numerical value of $\chi(1)$ from Fig.~\ref{fig:fig8}.

The asymptotic solution \eqref{eq:asymptotic-sols-appA}
for $\tau \gg b$ is shown in Fig.~\ref{fig:fig9} and
compares reasonably well with the numerical solution
in Fig.~\ref{fig:fig8}.
For completeness, the analytic pre-bounce solution
for $\tau \leq -b$ is also displayed in Fig.~\ref{fig:fig9}.
The mismatch at $\tau= \pm b$ in the left- and right-panels
of Fig.~\ref{fig:fig9} is of no concern,
as the asymptotic solution \eqref{eq:asymptotic-sols-appA}
is only approximative at $\tau=b$.

\end{appendix}

\end{document}